\documentclass[11pt,english]{article}
\usepackage[T1]{fontenc}
\usepackage[latin9]{inputenc}
\usepackage{amsmath}
\usepackage{graphicx}
\usepackage{amssymb}

\makeatletter

\usepackage{epsfig}\usepackage{graphics}

\newtheorem{theorem}{Theorem}

\newcommand{\E}{{\mathbb{E}}}
\newcommand{\Q}{{\mathbb{Q}}}

\def\Pr{{P}}
\def\Q{{\Bbb Q}}
\def\E{{\Bbb E}}

\def\LL{{\cal L}}
\def\0{{\bf 0}}

\def\b{\beta}

\def\phi{\varphi}

\def\T{\T}

\def\Cox{\hfill \Box}

\newcommand{\vectiiii}[4]{\ensuremath{\begin{pmatrix} #1 \\ #2 \\ #3 \\ #4 \end{pmatrix}}}

\catcode`@=11 \@addtoreset{equation}{section} \catcode`@=12

\usepackage{babel}

\begin{document}

\title{Multilayer Parking with Screening on a Random Tree}

\author{S.R.Fleurke%
\thanks{Agentschap Telecom, Postbus 450, 9700 AL Groningen, The Netherlands,
\texttt{sjoert.fleurke@at-ez.nl} %
},\, and C. Külske 
\thanks{Ruhr-Universität Bochum, Fakultät für Mathematik, Universitätsstraße
150, 44780 Bochum, Germany, \texttt{Christof.Kuelske@rub.de} %
} }

\maketitle

\begin{abstract}
In this paper we present a multilayer particle deposition model on a random tree.
We derive the time dependent densities of the first and second layer
analytically and show that in all trees the limiting density of the
first layer exceeds the density in the second layer. We also provide
a procedure to calculate higher layer densities and prove that random
trees have a higher limiting density in the first layer than regular
trees. Finally, we compare densities between the first and second
layer and between regular and random trees.
\end{abstract}
\smallskip{}

 \textit{Key--Words:} Car parking problem, Random sequential adsorption, Sequential frequency assignment process, Particle systems.

\section{Introduction}

Parking models with screening were first studied in the field of
ballistic particle deposition, see for example \cite{Meakin}. In
those models particles are moving towards a substrate or a fiber
until they encounter a previously deposited particle or the
substrate itself. A particle always tries to park on a layer as low
as possible but due to ``screening'' the particle cannot pass
formerly deposited particles. In our model the screening rule makes
every particle park in the highest layer possible where it is
supported by a particle in the layer below (see figure 1). This
``Tetris'' model is very different from the so-called Sequential
frequency assignment process (SFAP) in which particles (assignments)
can skip particles on their way down \cite{Dehling07}. In that model
particles are deposited in the lowest layer (frequency) possible. It
has been found for the SFAP that there is an increasing limiting
density of particles in higher layers due to boundary and other
effects \cite{Dehling07,Kulske}. In this paper we will show
analytically that in the model with screening the opposite is true.
The density in the first layer appears to be higher than in the
second layer. We conjecture that the same applies to the other
layers.

Also, we generalize the model on the one-dimensional lattice to a
model on regular and random trees. Recently, several random particle deposition
tree models have been studied in \cite{Dehling08, Gouet, Sudbury}.
However, to our knowledge this is the first time a multilayer random
tree model is treated.\\
\begin{figure}[htp]
\begin{centering}
\includegraphics[scale=0.45]{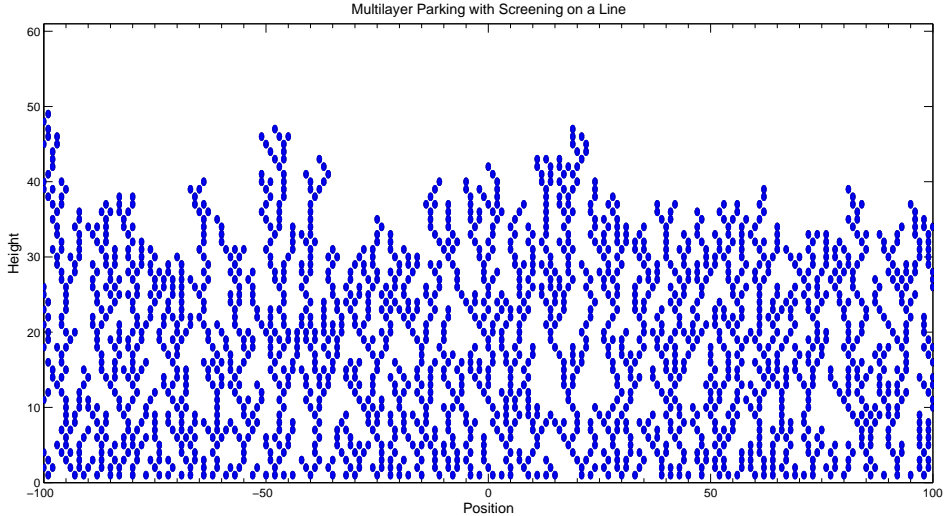}
\end{centering}
\caption{Example of a realization of the particle deposition process
with screening on a one-dimensional lattice. }
\label{example}
\end{figure}

\section{Layer Densities}

\subsection{The Dynamics}

The precise definition of the model is as follows. We consider a random
tree with vertices $i\in V$ and degree at the site $i$ given by
$D_{i}$. We choose $D_{i}$ to be independent random variables with
the same distribution $\mathbb{Q}$ given by \begin{equation}
\begin{split}\mathbb{Q}(D_{i}=k)=a_{k}\end{split}
\end{equation}
 on the integers starting from $2$. The latter requirement ensures
that we have no open ends with probability one.


We denote the generating function of the distribution by \begin{equation}
\begin{split}G(s)=\sum_{k=2}^{\infty}a_{k}s^{k}\end{split}
\end{equation}

We will denote the expected value with respect to this probability
distribution by the same symbol $\Q$. We fix a realization of the random tree,
and, denoting by $V$ its vertex set.

In the first part of the paper we will be interested in the behavior of
particle configurations arising from particle deposition
in the first two layers, that is we consider the marginal of an infinite layer particle
model on the first two layers.  This will be generalized to higher layers later.
To describe the behavior on the first two layers we consider (suitably coded) occupation
numbers $m=(m(i))_{i\in V}\in\Omega=\{0,1,2,3\}^{V}$.
Here the \textit{spin} $m(i)$ denotes the joint occupation numbers
at vertex $i$ at height $1$ and $2$. It is useful for short notation
to interpret the occupation numbers at various heights as binary digits
and write ordinary natural numbers. That is we write \begin{eqnarray}
m(i)=\left\{ \begin{array}{rl}
0 & \mbox{if vertex \ensuremath{i~}is vacant in the first and second line}\\
1 & \mbox{if vertex \ensuremath{i~}is occupied in the first but not in the second line}\\
2 & \mbox{if vertex \ensuremath{i~}is occupied in the second but not in the first line}\\
3 & \mbox{if vertex \ensuremath{i~}is occupied in the first and in the second line}\end{array}\right.\end{eqnarray}
 so that $m(i)\in\{0,1,2,3\}$.

Now we can describe the dynamics of the joint process of particle occupations
in the first two layers and the total number of particles which have arrived by the following
generator. Note that it is really necessary to consider also particle arrivals beyond
the first two layers because of
the screening effects higher layer particles might have on lower layers.
It is furthermore necessary to distinguish two sorts of particle arrivals: those
which change the lower layers and those which leave the lower layers unchanged.
Let $F$ be a joint function of particle occupations in the first two layers and particle numbers.
Then the generator of our process reads
\begin{equation}
\begin{split}\label{Generator}
\LL F(m,N) &=\sum_{k\in V}\Bigl( \sum_{s=1,2,3,4}F(m^{s,k},N^k)r_{k}(s;\mathfrak{M}_{k}) \Bigr.  \cr%
&+ \Bigl. \bigl(1- \sum_{s=1,2,3,4}r_{k}(s;\mathfrak{M}_{k})
\bigr)F(m,N^k) - F(m,N) \Bigr)
\end{split}
\end{equation}
with \begin{eqnarray}
N^{k}(i)=\left\{ \begin{array}{ll}
 N(k)+1& \mbox{if \ensuremath{k = i}}\\
N^{k}(i) & \mbox{if \ensuremath{k \neq i}}\end{array}\right.\end{eqnarray}
and with \begin{eqnarray}
m^{s,k}(i)=\left\{ \begin{array}{ll}
 s & \mbox{if \ensuremath{k=i}}\\
m(i) & \mbox{if \ensuremath{k \neq i}}\end{array}\right.\end{eqnarray}
where $\mathfrak{M}_{k}:=(N(l),m(l))_{l\in\{k\}\cup C(k)}$ where $C(k):=\left\{ i:dist(i,k)=1\right\} $
is the neighborhood of vertex $k$.

So, the first term in the generator
describes the events when the addition of a new particle also changes the configuration
in one of the first two layers. The second term in the generator describes the events
when the first two layers are already full or screened, and a further adding of a particle
does not change its filling.

The rates are either equal to zero or one. They are $1$ precisely
in the following cases (see also Figure \ref{fig:01}) listed below.
\begin{enumerate}
\item {$0\mapsto1$} Adding a particle in the first line at vertex $i$. We
have\begin{equation}
\begin{split}r(1;\{N(i)=0\}\cap\{\forall_{j\in C(i)}:N(j)=0\})=1\end{split}
\end{equation}
 Indeed, this occurs when the site and all its neighbors are empty
in all layers.
\item {$0\mapsto2$} Adding a particle in the second line at $i$ while the
first line was empty at the site
\begin{equation}
\begin{split}r(2;\{N(i)=0, \exists J\subset C(i) : \forall j \in J: m(j)=1,N(j)=1, \forall k\in C(i)\backslash J :\\
 N(k)=0\})=1\end{split}
\end{equation}

 It is only possible to reach the state $m(i)=2$ when at least one
of the neighbors has a particle at layer 1 and all the others are
totally empty.
\item {$1\mapsto3$} Adding a particle in the second line while the first line
was full at the site \begin{equation}
\begin{split}
r(3;\{m(i)=1,N(i)=1\}\cap\{\forall_{j\in C(i)}:
N(j)=0\})=1\end{split}
\end{equation}
 To get into state $m(i)=3$ there must be one particle in vertex
$i$ and all neighboring sites should be empty to avoid screening.\\

All other transitions are impossible.%
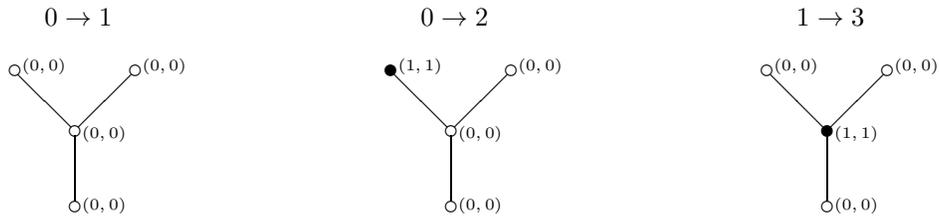
\begin{figure}[h]
 \setlength{\unitlength}{1mm} \begin{picture}(150,35)(0,-7)\put(21,24){\small {$0\rightarrow1$}}
\put(71,24){\small {$0\rightarrow2$}} \put(121,24){\small {$1\rightarrow3$}}\multiput(25,0.6)(50,0){3}{\line(0,1){8.8}}
\multiput(24.5,10.5)(50,0){3}{\line(-1,1){7}} \multiput(25.5,10.5)(50,0){3}{\line(1,1){7}}
\multiput(25,10)(50,0){3}{\circle{1.5}} \multiput(25,-0.1)(50,0){3}{\circle{1.5}}
\multiput(17,18)(50,0){3}{\circle{1.5}} \multiput(33,18)(50,0){3}{\circle{1.5}}
\put(67,18){\circle*{1.5}} \put(125,10){\circle*{1.5}}
\put(26,-0.5){\tiny {$(0,0)$}} \put(76,-0.5){\tiny {$(0,0)$}}
\put(126,-0.5){\tiny {$(0,0)$}} \put(26,9){\tiny {$(0,0)$}}
\put(76,9){\tiny {$(0,0)$}} \put(126,9){\tiny {$(1,1)$}} \put(18,18){\tiny {$(0,0)$}}
\put(68,18){\tiny {$(1,1)$}} \put(118,18){\tiny {$(0,0)$}}
\put(34,18){\tiny {$(0,0)$}} \put(84,18){\tiny {$(0,0)$}}
\put(134,18){\tiny {$(0,0)$}}\end{picture}

\caption{Neighborhood configurations of a vertex that allow the $m_{t}$ transitions
$0\rightarrow1$, $1\rightarrow2$ and $1\rightarrow3$ respectively.
In this example the central vertex has three neighbors. The states
are denoted with the notation $(N_{t},m_{t})$, e.g. $(1,1)$ means
that one particle arrived and that it was deposited on the first layer.
In order to have a transition from $0$ to $1$ every vertex in the
neighborhood has to be totally empty ($N_{t}=0$). To get a transition
from $0$ to $2$ the vertex itself must be empty, but at least one
of the neighbors has to have one particle in total that lies on the
first layer. Finally, to have a transition $1$ to $3$ the vertex
has to have exactly one particle located on the first layer, while
the neighbors should be empty.}

\label{fig:01}
\end{figure}

\end{enumerate}

This generator defines a time-homogeneous Markov
jump process on the infinite graph by standard theory \cite{Liggett}
such that (\ref{Generator})
$\frac{d}{dt}|_{t=0}\E^{m,N} F(m_t,N_t)=\LL F(m,N)$.

Here $\E^{m,N}$ denotes the expected value with respect to the process,
started in the initial configuration $(m,N)=(m(i),N(i))_{i\in V}$ at
$t=0$.

We underline that we consider the marginal of the first two layers of
a model where particles may pile up to arbitrary high layers. The
present model differs from the model discussed in \cite{Kulske} where
particles that cannot be deposited in the first or second layer are
rejected.

\subsection{Regular Trees}

We first consider the densities, taken at an arbitrary vertex called $0$,
\begin{equation}
\begin{split}
\rho_{t}^{d}(1)&=\Pr_{t} ( m(0)=1 )+\Pr_{t} ( m(0)=3 )\cr
\rho_{t}^{d}(2)&=\Pr_{t} ( m(0)=2 )+\Pr_{t} ( m(0)=3 )\cr
\end{split}
\end{equation}
on the first and second layer on a
regular tree with degree $d\ge2$. Having understood their behavior on a regular tree we can derive
the densities on random trees easily in section \ref{sect:randomtrees}.

\begin{theorem} Consider the regular tree $T_{d}$ with degree $d\ge2$.
Particles arrive at the vertices of $T_{d}$ according to a Poisson
process and obey the screening rules of deposition. Then the time
dependent densities are, on the first layer \begin{eqnarray}
\rho_{t}^{d}(1)=\frac{1 - e^{-(d+1)t}}{d+1}\end{eqnarray}
 and on the second layer \begin{eqnarray}
\rho_{t}^{d}(2) & = & \left(\frac{d}{d-1}\right)^{d}\sum_{k=0}^{d}{d \choose k}\frac{d^{-k}(-1)^{k}}{(d-1)k+d+1}-\frac{d}{(d+1)^{2}}+\frac{d}{(d+1)^{2}}e^{-(d+1)t}\nonumber \\
 & - & \left(\frac{d}{d-1}\right)^{d}\sum_{k=0}^{d}{d \choose k}\frac{d^{-k}(-1)^{k}}{(d-1)k+d+1}e^{-[(d-1)k+d+1]t}-\frac{1}{d+1}te^{-(d+1)t}~~~~\end{eqnarray}
\label{theorem:regtrees} \end{theorem}

\textbf{Proof of Theorem \ref{theorem:regtrees}:}\\
Throughout the paper we use the notation
 $$D_{t}^{d}(s)=
\Pr_{t} ( m(0)=s )$$ for all $s$.
 We fix at a certain vertex $0$.
  The surrounding vertices are
numbered $1,2,\dots,d$. First we calculate the time derivative of  $D_{t}^{d}(1)$
and integrate back. Taking into account the first and third process depicted
in Fig. 2 we see that
\begin{eqnarray}
\dot{D}_{t}^{d}(1) & = & \Pr_{t}(N_{t}(0)=0,\forall_{k\le d}N_{t}(k)=0)-\Pr_{t}(N_{t}(0)=1,\forall_{k\le d}N_{t}(k)=0) \nonumber\\
 & = & e^{-(d+1)t}-te^{-t}e^{-dt} \nonumber\\
 & = & -(t-1)e^{-(d+1)t}\end{eqnarray}
 So, now with $D_{0}^{d}(1)=0$ we find \begin{eqnarray}
D_{t}^{d}(1) & = & \frac{d}{(d+1)^{2}}-\frac{d}{(d+1)^{2}}e^{-(d+1)t}+\frac{1}{d+1}te^{-(d+1)t}\end{eqnarray}
 We apply the same technique to find $D_{t}^{d}(2)$ \begin{eqnarray}
\dot{D}_{t}^{d}(2) & = & \sum_{k=1}^{d}{d \choose k}\Pr_{t}(N_{t}(0)=0,\forall_{i\le k}N_{t}(i)=1,m_{t}(i)=1,\forall_{k<j\le d}N_{t}(j)=0) \nonumber\\
 & = & \sum_{k=1}^{d}{d \choose k}\Pr_{t}(\forall_{i\le k}N_{t}(i)=1,m_{t}(i)=1,\forall_{k<j\le d}N_{t}(j)=0|N_{t}(0)=0)e^{-t}~~~~~~~\nonumber \\
 & = & e^{-t}\sum_{k=1}^{d}{d \choose k}[\Pr_{t}(N_{t}(1)=1,m_{t}(1)=1|N_{t}(0)=0)]^{k}[\Pr_{t}(N_{t}(0)=0)]^{d-k}\end{eqnarray}
 Now we need to calculate the quantity $S_{t}^{d}:=\Pr_{t}^{d}(N_{t}(1)=1,m_{t}(1)=1|N_{t}(0)=0)$.
This is done by constructing another differential equation.\begin{eqnarray}
\dot{S}_{t}^{d} & = & \Pr_{t}(N_{t}(1)=0,\forall_{i:d(i,1)=1}N_{t}(i)=0|N_{t}(0)=0)-S_{t}^{d} \nonumber \\
 & = & e^{-dt}-S_{t}^{d}\end{eqnarray} whose solution is
\begin{eqnarray}
S_{t}^{d}= e^{-t} \frac{1 - e^{-(d-1)t}}{d-1} \label{S^d} %
\end{eqnarray}
 So, we have\begin{eqnarray}
\dot{D}_{t}^{d}(2) & = & e^{-t}\sum_{k=1}^{d}{d \choose k}\left[\frac{1}{d-1}e^{-t}(1-e^{-(d-1)t})\right]^{k}[e^{-t}]^{d-k} \nonumber \\
 & = & e^{-(d+1)t}\sum_{k=1}^{d}{d \choose k}\left[\frac{1-e^{-(d-1)t}}{d-1}\right]^{k} \nonumber \\
 & = & e^{-(d+1)t}\left[\left(1+\frac{1-e^{-(d-1)t}}{d-1}\right)^{d}-1\right] \nonumber \\
 & = & -e^{-(d+1)t}+e^{-(d+1)t}\left(\frac{d}{d-1}-\frac{1}{d-1}e^{-(d-1)t}\right)^{d}\nonumber \\
 & = & -e^{-(d+1)t}+e^{-(d+1)t}\sum_{k=0}^{d}{d \choose k}\left(\frac{d}{d-1}\right)^{d-k}\left(\frac{-1}{d-1}\right)^{k}e^{-(d-1)kt}~~\nonumber\\
 & = & -e^{-(d+1)t}+\left(\frac{d}{d-1}\right)^{d}\sum_{k=0}^{d}{d \choose k}d^{-k}(-1)^{k}e^{-[(d-1)k+d+1]t}\end{eqnarray}
 which gives (with $D_{0}^{d}(2)=0$) \begin{eqnarray}
D_{t}^{d}(2) & = & \left(\frac{d}{d-1}\right)^{d}\sum_{k=0}^{d}{d \choose k}\frac{d^{-k}(-1)^{k}}{(d-1)k+d+1}-\frac{1}{d+1}\nonumber \\
 &  & +\frac{1}{d+1}e^{-(d+1)t}-\left(\frac{d}{d-1}\right)^{d}\sum_{k=0}^{d}{d \choose k}\frac{d^{-k}(-1)^{k}}{(d-1)k+d+1}e^{-[(d-1)k+d+1]t}~~~~~~~~\end{eqnarray}
 Finally, for $D_{t}^{d}(3)$ we find \begin{eqnarray}
\dot{D}_{t}^{d}(3) & = & \Pr_{t}(N_{t}(0)=1,\forall_{k\le d}N_{t}(k)=0) \nonumber\\
 & = & te^{-(d+1)t} \end{eqnarray}
 so, that (with $D_{0}^{d}(3)=0$) \begin{eqnarray}
D_{t}^{d}(3) & = & \frac{1}{(d+1)^{2}}-\frac{1}{(d+1)^{2}}e^{-(d+1)t}-\frac{1}{d+1}te^{-(d+1)t}\end{eqnarray}
 The densities of the first and second layer follow immediately by
adding $D_{t}^{d}(1)$ and $D_{t}^{d}(3)$ for the first layer, and
$D_{t}^{d}(2)$ and $D_{t}^{d}(3)$ for the second layer. $\Cox$
\\
\\
{\bf Remark:} Note that for the derivation of the formula for the
first layer we did not have to use the absence of loops
in a tree. Therefore, the first layer density on a graph is the same
as on a tree, no matter whether they are regular or random.\\

\subsection{Random Trees}

\label{sect:randomtrees} Let us now consider the case of particle deposition
on a random tree where the number of neighbors of every vertex is
a random number according to some $G(s)=\sum_{n=2}^{\infty}a_{n}s^{n}$.
We now have the following

\begin{theorem} \label{theorem:randomtrees} Consider a multilayer
random tree $T_{D}$ with generating function \\
 $G_T(s)=\sum_{n=2}^{\infty}a_{n}s^{n}$. Particles arrive at the
vertices of $T_{D}$ according to a Poisson process and obey the
screening rules of deposition. Then the tree-averaged time dependent
densities are, on the first layer \begin{eqnarray}
\mathbb{Q}\rho_{t}(1) & = &
\sum_{k=2}^{\infty}a_{k}\frac{\left( 1-e^{-(k+1)t} \right)}{k+1}\end{eqnarray}%
 and on the second layer \begin{eqnarray}
\mathbb{Q}\rho_{t}(2) & = & \sum_{k=2}^{\infty}\frac{a_{k}}{(k+1)^{2}}-\sum_{k=2}^{\infty}\left(\frac{a_{k}}{(k+1)^{2}}e^{-(k+1)t}+\frac{a_{k}}{k+1}te^{-(k+1)t}\right)\\
 &  & +\sum_{d_{0}=2}^{\infty}a_{d_{0}}\sum_{k=1}^{d_{0}}{d_{0} \choose k}\sum_{i=0}^{k}{k \choose i}(-1)^{i}\left(\sum_{d=2}^{\infty}\frac{a_{d}}{d-1}\right)^{k-i}\int_{0}^{^{t}}Z_{u}^{^{i}}e^{-(d_{0}+1)u}du~~~~~~~~~
 \end{eqnarray}
where
$Z_{t}:=\sum_{d=2}^{\infty}a_{d}e^{-(d-1)t}/(d-1)
$.

\end{theorem} \textbf{Proof:}\\
 First, we calculate $\mathbb{Q}D_{t}(1)$ and $\mathbb{Q}D_{t}(3)$. Notice
that the derivatives of these functions in a certain vertex $0$ are
not affected by the tree ensemble beyond the nearest neighbors. Therefore,
we can immediately start averaging $D_{t}(1)$ and $D_{t}(3)$ over
$\mathbb{Q}$ rather than dealing with its derivatives first. In the
previous section we already found \begin{eqnarray}
D_{t}^{d}(1) & = & \frac{d}{(d+1)^{2}}-\frac{d}{(d+1)^{2}}e^{-(d+1)t}+\frac{1}{d+1}te^{-(d+1)t}\end{eqnarray}
where $d$ now denotes the (random)
number of nearest neighbors of the site under consideration.
 Averaging over $\mathbb{Q}$ results then  in \begin{eqnarray}
\mathbb{Q}D_{t}(1) & = & \sum_{k=2}^{\infty}\frac{a_{k}k}{(k+1)^{2}}-\sum_{k=2}^{\infty}\left(\frac{a_{k}k}{(k+1)^{2}}e^{-(k+1)t}-\frac{a_{k}}{k+1}te^{-(k+1)t}\right)\end{eqnarray}
 Similarly, we find \begin{eqnarray}
\mathbb{Q}D_{t}(3) & = & \sum_{k=2}^{\infty}\frac{a_{k}}{(k+1)^{2}}-\sum_{k=2}^{\infty}\left(\frac{a_{k}}{(k+1)^{2}}e^{-(k+1)t}+\frac{a_{k}}{k+1}te^{-(k+1)t}\right)\end{eqnarray}
 Adding these two results gives the density on the first layer. \\
 In the previous section we already found \begin{eqnarray}
\dot{D}_{t}(2) &=& e^{-t}\sum_{k=1}^{d_{0}}{d_{0} \choose
k}[\Pr_{t}^{d_{k}}(N_{t}(k)=1,m_{t}(1)=1|N_{t}(0)=0)]^{k}[\Pr_{t}(N_{t}(0)=0)]^{d_{0}-k}~~~~~~~~\end{eqnarray}
 where the number of neighbors of vertex $i$ is denoted by $d_{i}$.
Note that in this section the $d_{i}$'s may be different since we
are treating a random tree. So, we get\\
 \begin{eqnarray}
\dot{D}_{t}(2) & = & e^{-t}\sum_{k=1}^{d_{0}}{d_{0} \choose k}[S_{t}^{d_{k}}]^{k}[\Pr_{t}(N_{t}(0)=0)]^{d_{0}-k}\Rightarrow\\
\mathbb{Q}\dot{D}_{t}(2) & = &
e^{-t}\sum_{d_{0}=2}^{\infty}a_{d_{0}}\sum_{k=1}^{d_{0}}{d_{0}
\choose
k}\left[\sum_{d_{k}=2}^{\infty}a_{d_k}S_{t}^{d_{k}}\right]^{k}[\Pr_{t}(N_{t}(0)=0)]^{d_{0}-k}\end{eqnarray}
In (\ref{S^d}) we already found that
$S_{t}^{d_{k}}=\frac{1}{d_{k}-1}e^{-t}(1- e^{-(d_{k}-1)t})$. So, we
have
\begin{eqnarray}
\mathbb{Q}\dot{D_{t}}(2) & = & \sum_{d_{0}=2}^{\infty}a_{d_{0}}e^{-(d_{0}+1)t}\sum_{k=1}^{d_{0}}{d_{0} \choose k}\left[\sum_{d=2}^{\infty}a_{d}\frac{1}{d-1}-\sum_{d=2}^{\infty}a_{d}\frac{e^{-(d-1)t}}{d-1}\right]^{k} \nonumber \\
 & = & \sum_{d_{0}=2}^{\infty}a_{d_{0}}\sum_{k=1}^{d_{0}}{d_{0} \choose k}\sum_{i=0}^{k}{k \choose i}(-1)^{i}\left(\sum_{d=2}^{\infty}a_{d}\frac{1}{d-1}\right)^{k-i}\nonumber \\
 & \times & \left(\sum_{d=2}^{\infty}a_{d}\frac{e^{-(d-1)t}}{d-1}\right)^{i}e^{-(d_{0}+1)t} \nonumber\\
 & = & \sum_{d_{0}=2}^{\infty}a_{d_{0}}\sum_{k=1}^{d_{0}}{d_{0} \choose k}\sum_{i=0}^{k}{k \choose i}(-1)^{i}\left(\sum_{d=2}^{\infty}\frac{a_{d}}{d-1}\right)^{k-i}Z_{t}^{^{i}}e^{-(d_{0}+1)t}
\end{eqnarray}
with
$Z_{t}:=\sum_{d=2}^{\infty}a_{d}e^{-(d-1)t}/(d-1)$. 
So, by integration we find
\begin{eqnarray}
\mathbb{Q}D_{t}(2) & = &
\sum_{d_{0}=2}^{\infty}a_{d_{0}}\sum_{k=1}^{d_{0}}{d_{0} \choose
k}\sum_{i=0}^{k}{k \choose
i}(-1)^{i}\left(\sum_{d=2}^{\infty}a_{d}\frac{1}{d-1}\right)^{k-i}\int_{0}^{^{t}}Z_{u}^{^{i}}e^{-(d_{0}+1)u}du~~~~~~~~~\end{eqnarray}
$\Cox$

\subsubsection*{Example}

We would like to give an example where a closed-form solution is available
which is free of integrals and gives us the time-dependent behavior of
densities in the first and second line as sums whose main terms are exponentials in the time.
Let us consider the
special case where there are only two possible numbers of neighbors
$a$ and $b$ on the random tree, i.e. we take $G(s)=p_{a}s^{a}+p_{b}s^{b}$.  We find
\begin{eqnarray}
\mathbb{Q}\rho_{t}(1) & = & \frac{p_{a}}{a+1}\left( 1 -
e^{-(a+1)t}\right) + \frac{p_{b}}{b+1} \left(1- e^{-(b+1)t} \right)
\end{eqnarray}
 For the second layer we need to calculate the quantity $C_t(n, x) := \int_{0}^{^{t}}Z_{u}^{^{n}}e^{-(x+1)u}du$.
We have
\begin{eqnarray} %
Z_{t}^{n} & = & \left[ \frac{p_a}{a-1} e^{-(a-1)t} + \frac{p_b}{b-1} e^{-(b-1)t} \right]^n \nonumber \\%
& = & \sum_{j=0}^{n}{n \choose j} \left(\frac{p_{a}}{a-1}\right)^{n-j} \left(\frac{p_{b}}{b-1}\right)^{j}e^{-[(a-1)(n-j)+(b-1)j]t} %
\end{eqnarray}
 and so
\begin{eqnarray}
C_t(n,x)=\left\{ \begin{array}{rl}
\frac{1}{x+1}\left(1-e^{-(x+1)t} \right) & \mbox{if } n=0 \\
\sum_{j=0}^{n}{n \choose j}\frac{\left(\frac{p_{a}}{a-1}\right)^{n-j}\left(\frac{p_{b}}{b-1}\right)^{j}\left( 1 - e^{-[(a-1)(n-j)+(b-1)j+x+1]t} \right)}{(a-1)(n-j)+(b-1)j+x+1}~~%
 & \mbox{if } n>0
\end{array}\right.
\end{eqnarray}
So, for the second layers density we find the closed form
\begin{eqnarray}
\mathbb{Q}\rho_{t}(2)
\!\!\! &=& \!\!\!p_{a} \left( \frac{1}{(a+1)^2} -\frac{e^{-(a+1)t}}{(a+1)^2}- \frac{te^{-(a+1)t}}{a+1}\right) + p_{b} \left( \frac{1}{(b+1)^2} -\frac{e^{-(b+1)t}}{(b+1)^2}- \frac{te^{-(b+1)t}}{b+1}\right) \nonumber \\
 & + & p_{a}\sum_{k=1}^{a}{a \choose k}\sum_{i=0}^{k}{k \choose i}(-1)^{i}\left(\frac{p_{a}}{a-1}+\frac{p_{b}}{b-1}\right)^{k-i} C_t(i,a)~~~~~~ \nonumber \\%
 & + & p_{b}\sum_{k=1}^{b}{b \choose k}\sum_{i=0}^{k}{k \choose i}(-1)^{i}\left(\frac{p_{a}}{a-1}+\frac{p_{b}}{b-1}\right)^{k-i} C_t(i,b)~~~~~~%
\end{eqnarray}

\subsection{Procedure to Derive Higher Layer Densities}

It is natural to ask whether the procedure we just described
to obtain densities on the first two layers
can be generalized to obtain densities in a finite number of layers.
To see the issue of higher layers more clearly let us specialize from the tree
to the line. In this case, we claim that the time-dependent
probabilities of the occurrence of any single-site pattern describing occupations
up to a given finite height can in principle be calculated.
However, in most cases the (probability of occurence of a) pattern
can not be calculated directly but by a recursive algorithm
which involves the computation of simpler patterns which we call the pre-image motives.

The following procedure provides a method to find the time-dependent
formula of the proportion of any pattern on a vertex. It consists
of four steps: 1. find the pre-image motives (the configurations from
which the pattern under interest can increase or decrease), 2. obtain the solutions
of probabilities for occurence of the pre-image motives, 3. construct a differential equation of
the target pattern based on the pre-image motives, and finally 4.
solve the differential equation.
\\
As an example how the program works we will now calculate the probability of the occurrence
of $Y_{t}=(0,1,0,1)_{t}'$, meaning the probability that the first
and third layer are occupied and the second and fourth layer are empty.
That the procedure stops after finitely many steps is not obvious from the beginning.
Responsible for this fact is the screening. This will become clear in the example below.
In a model without screening like  \cite{Kulske} it is not true, and a corresponding
recursion produces an infinite number of local motives.

\subsubsection{Step 1: Find the pre-image motives}

In this step we have to find the patterns whose occurrences contribute
to an increase or decrease of our target pattern. In the case of $Y_{t}=(0,1,0,1)'_{t}$
we find four pre-image motives, i.e. \begin{eqnarray}
A_{1}=\begin{pmatrix}\times & \times & \times\\
1 & 0 & \times\\
0 & 1 & 0\end{pmatrix}\!\!,A_{2}=\begin{pmatrix}\times & \times & \times\\
1 & 0 & 1\\
0 & 1 & 0\end{pmatrix}\!\!,A_{3}=\begin{pmatrix}\times & \times & \times\\
0 & 1 & 0\\
1 & 0 & 0\\
0 & 1 & 0\end{pmatrix}\!\!,A_{4}=\begin{pmatrix}\times & \times & \times\\
0 & 1 & 0\\
1 & 0 & 1\\
0 & 1 & 0\end{pmatrix}~\end{eqnarray}
 This notation indicates on which position and layer a particle has
been deposited ($1$)  and where not ($0$), and where no particles have arrived so
far ($\times$). A particle is denoted by a $1$ and empty positions that will
remain empty due to blocking of neighbors or to the screening effect
are indicated with a $0$. Positions at and beyond which no particle
has arrived so far are indicated with a $\times$. Indeed, the
proportion of the occurrence of $Y_{t}$ will increase with the
proportion of both $A_{1}$ and $A_{2}$. In both patterns a particle
is able to be deposited on the third layer and complete the pattern
of $Y_{t}$. The new particle can not be screened by particles in
higher layers. On the other hand, the occurrence of $A_{3}$ or
$A_{4}$ may lead to a decrease of $Y_{t}$, because they allow the
arrival of a particle in the center location which results in
$(1,1,0,1)'$. There are no other motives that can directly influence
the proportion of $Y_{t}$.

\subsubsection{Step 2: Obtain the solutions of the pre-image motives}

In this step we treat the pre-image motives one-by-one and find their
solutions using the same four step procedure again. First we look at $A_{1}$
and detect its pre-image motives.

\subsubsection*{Finding $A_{1}$}

We apply the same procedure to find $A_{1}$. With an abuse of notation
we write $A_1(t)$ for the probabiity of its occurence.

\textbf{Step 1':} The pre-image motives of $A_{1}$ are:\\
 \begin{eqnarray}
B_{1}=\begin{pmatrix}\times & \times & \times & \times\\
\times & \times & \times & \times\\
\times & 0 & 1 & 0\end{pmatrix},B_{2}=\begin{pmatrix}\times & \times & \times & \times\\
\times & \times & \times & \times\\
1 & 0 & 1 & 0\end{pmatrix},A_{1}=\begin{pmatrix}\times & \times & \times\\
1 & 0 & \times\\
0 & 1 & 0\end{pmatrix}\end{eqnarray}

\textbf{Step 2':} Solutions of the pre-image motives of $A_{1}$\\
 \begin{eqnarray}
B_{1}(t)=te^{-4t}\end{eqnarray}
 \begin{eqnarray}
B_{2}(t) & = & te^{-3t}\Pr_{t}(N_{t}(1)=0,m_{t}(1)=1|N_{t}(0)=0) \nonumber\\
 & = & S_{t}^{2}te^{-3t} \nonumber\\
 & = & te^{-4t}-te^{-5t}\end{eqnarray}
 where we used our earlier result for $S_{t}^{d}$ in (\ref{S^d}).

\textbf{Step 3':} The differential equation for $A_{1}(t)$ takes the form \\
 \begin{eqnarray}
\dot{A}_{1}(t) & = & B_{1}(t)+B_{2}(t)-3A_{1}(t)\nonumber \\
 & = & 2te^{-4t}-te^{-5t}-3A_{1}(t)\end{eqnarray}

\textbf{Step 4':} Solution of $A_{1}(t)$ \\

Together with $A_{1}(0)=0$ we find
\begin{eqnarray}
A_{1}(t)=\frac{7}{4}e^{-3t}-2e^{-4t}-2te^{-4t}+\frac{1}{4}e^{-5t}+\frac{1}{2}te^{-5t}
\end{eqnarray}

\subsubsection*{Finding $A_{2}$}

We apply the same steps to get $A_{2}$.\\

\textbf{Step 1':} The pre-image motives of $A_{2}$ are \begin{eqnarray}
C_{1}(t)=\begin{pmatrix}\times & \times & \times & \times\\
\times & \times & 0 & 1\\
\times & 0 & 1 & 0\end{pmatrix},C_{2}(t)=\begin{pmatrix}\times & \times & \times & \times\\
\times & \times & 0 & 1\\
1 & 0 & 1 & 0\end{pmatrix},A_{2}(t)=\begin{pmatrix}\times & \times & \times\\
1 & 0 & 1\\
0 & 1 & 0\end{pmatrix}\end{eqnarray}

\textbf{Step 2':} Now we solve $C_{1}(t)$ and $C_{2}(t)$.\\
 We find $C_{1}(t)=\begin{pmatrix}\times & \times & \times & \times\\
\times & \times & 0 & 1\\
\times & 0 & 1 & 0\end{pmatrix}=U_{t}e^{-2t}$. With $U_{t}:=\Pr_{t}(N_{t}(1)=1,m_{t}(1)=1,N_{t}(2)=1,m_{t}(2)=2|N_{t}(0)=0)$.
Now, we have to solve $U_{t}$ first. \begin{eqnarray}
\dot{U}_{t} & = & -2U_{t}+\Pr_{t}(N_{t}(1)=1,N_{t}(2)=0,N_{t}(3)=0|N_{t}(0)=0)\nonumber\\
 &  & +\Pr_{t}(N_{t}(1)=1,N_{t}(2)=0,N_{t}(3)=1,m_{t}(3)=1|N_{t}(0)=0)\nonumber\\
 & = & -2U_{t}+te^{-3t}\nonumber \\
 & + & \Pr_{t}(N_{t}(1)=1,N_{t}(3)=1,m_{t}(3)=1|N_{t}(0)=0,N_{t}(2)=0)e^{-t}\nonumber\\
 & = & -2U_{t}+te^{-3t}+\Pr_{t}(N_{t}(1)=1|N_{t}(0)=0,N_{t}(2)=0)\nonumber \\
 &  & \times\Pr_{t}(N_{t}(1)=1,m_{t}(1)=1|N_{t}(0)=0)e^{-t}\nonumber\\
 & = & -2U_{t}+te^{-3t}+te^{-t}S_{t}^{2}e^{-t}\nonumber\\
 & = & -2U_{t}+te^{-3t}+te^{-2t}(e^{-t}-e^{-2t})\nonumber\\
 & = & -2U_{t}+2te^{-3t}-te^{-4t}\end{eqnarray}
 So, we have to solve $\dot{U}_{t}+2U_{t}=2te^{-3t}-te^{-4t}$. The homogeneous
solution is $U_{t,hom}=Ce^{-2t}$, and a particular solution is
$U_{t,part}=-2te^{-3t}-2e^{-3t}+\frac{1}{2}te^{-4t}+\frac{1}{4}e^{-4t}$.
For $t=0$ we have $U_{0}=0$. So, this gives the general solution
$U_{t}=\frac{7}{4}e^{-2t}-2te^{-3t}-2e^{-3t}+\frac{1}{2}te^{-4t}+\frac{1}{4}e^{-4t}$.
Therefore, we have
$C_{1}(t)=U_{t}e^{-2t}=1\frac{3}{4}e^{-4t}-2te^{-5t}-2e^{-5t}+\frac{1}{2}te^{-6t}+\frac{1}{4}e^{-6t}$.
Now, we treat $C_{2}(t)$. We find \begin{eqnarray}
C_{2}(t) & = & \begin{pmatrix}\times & \times & \times & \times\\
\times & \times & 0 & 1\\
1 & 0 & 1 & 0\end{pmatrix}=U_{t}S_{t}^{2}e^{-t}\end{eqnarray}
 So, \begin{eqnarray}
C_{2}(t)=\left(1\frac{3}{4}e^{-2t}-2te^{-3t}-2e^{-3t}+\frac{1}{2}te^{-4t}+\frac{1}{4}e^{-4t}\right)(e^{-t}-e^{-2t})e^{-t}\end{eqnarray}

\textbf{Step 3':} \begin{eqnarray}
\dot{A}_{2}(t) & = & 2C_{1}(t)+2C_{2}(t)-3A_{2}(t)\nonumber \\
 & = & 7e^{-4t}-11\frac{1}{2}e^{-5t}-8te^{-5t}+5e^{-6t}+6te^{-6t}-\frac{1}{2}e^{-7t}-te^{-7t}-3A_{2}(t)~~~~\end{eqnarray}

\textbf{Step 4':}

So $A_{2}^{h}(t)=Ce^{-3t}$ and
$A_{2}^{p}(t)=-7e^{-4t}+\frac{31}{4}e^{-5t}+4te^{-5t}-\frac{7}{3}e^{-6t}-2te^{-6t}+\frac{3}{16}e^{-7t}+\frac{1}{4}te^{-7t}$
and the general solution becomes
\begin{eqnarray}
A_{2}(t)=\frac{67}{48}e^{-3t}-7e^{-4t}+\frac{31}{4}e^{-5t}+4te^{-5t}-\frac{7}{3}e^{-6t}-2te^{-6t}+\frac{3}{16}e^{-7t}+\frac{1}{4}te^{-7t}
\end{eqnarray}

\subsubsection*{Finding $A_{3}$ and $A_{4}$}

The last two motives are much less complicated compared with the former
two, so we can treat them together in one time.\\

\textbf{Step 1':} The motives of $A_{3}$ are $A_{1}$ and $A_{3}$
itself, whereas the motives of $A_{4}$ are $A_{2}$ and $A_{4}$.

\textbf{Step 2':} We already solved $A_{1}$ and $A_{2}$ above.

\textbf{Step 3':} The differential equations that we need to solve
are\begin{eqnarray} \dot{A}_{3}(t) =  A(t)-3A_{3}(t)\mbox{ and
}\dot{A}_{4}(t)=A_{2}(t)-3A_{4}(t)\end{eqnarray}
 respectively.

\textbf{Step 4':} The reader is invited to check that the solutions
are
\begin{eqnarray}
A_{3}(t) & = & -\frac{15}{4}e^{-3t}+\frac{7}{4}te^{-3t}+4e^{-4t}+2te^{-4t}-\frac{1}{4}e^{-5t}-\frac{1}{4}te^{-5t}\mbox{ and}\\
A_{4}(t) & = & -\frac{49}{16}e^{-3t}+\frac{67}{48}te^{-3t}+7e^{-4t}-\frac{39}{8}e^{-5t}-2te^{-5t}\nonumber \\
 & + & e^{-6t}+\frac{2}{3}te^{-6t}-\frac{1}{16}e^{-7t}-\frac{1}{16}te^{-7t}\end{eqnarray}

\subsubsection{Step 3: Construct a differential equation}

The differential equation for $Y_{t}$ is \begin{eqnarray}
\dot{Y}_{t} & = & 2A_{1}(t)+A_{2}(t)-2A_{3}(t)-A_{4}(t)\end{eqnarray}
 Indeed, the appearance of $Y_{t}$ can increase by $A_{1}(t)$ and
by its mirror motive. So, it counts two times. Also $A_{2}(t)$ increases
the proportion of $Y_{t}$ but only one time, because its pattern
is symmetric. Decrease of $Y_{t}$ occurs when a particle parks on
top of $A_{3}(t)$ or $A_{4}(t)$ where the former counts two times
because its mirror pattern has the same effect.

\subsubsection{Step 4: Solve the differential equation}

After some calculations we find, using $Y_{0}=0$: \begin{eqnarray}
\Pr_{t}\vectiiii{0}{1}{0}{1} & = & \frac{34}{735}-\frac{1991}{432}e^{-3t}+\frac{235}{144}te^{-3t}+7e^{-4t}+2te^{-4t}-\frac{121}{40}e^{-5t}-\frac{3}{2}te^{-5t}~~~~\nonumber \\
 & + & \frac{17}{27}e^{-6t}+\frac{4}{9}te^{-6t}-\frac{33}{784}e^{-7t}-\frac{5}{112}te^{-7t}\end{eqnarray}
 %

%
%
%
%
%

It should be clear that the probabilities of any other pattern
can be computed in a similar way.

\subsection{Comparison Results}

Let us now come back to the behavior on the first two layers and conclude
the paper with a discussion of comparison statements for densities.

\begin{theorem} Consider a regular tree $T_{d}$. The limiting density
in the first layer is higher than in the second layer for all $d\ge2$.
\end{theorem}

\textbf{Proof:} Let us denote $\b_{k}(d):=(d/(d-1))^{d}{d \choose
k}\frac{d^{-k}}{(d-1)k+d+1}$, so $\rho_{\infty}^{d}(2) =
\sum_{k=0}^{d} (-1)^{k}\b_{k}(d)-\frac{d}{(d+1)^{2}}$. One verifies
that $\b_{k+1}(d)/\b_{k}(d)<1$ so that $k\mapsto\b_{k}(d)$ is
decreasing. Therefore, making use of the alternating nature of the
sum, we have the bound
$\lim_{t\rightarrow\infty}\rho_{t}^{d}(2)<\b_{0}(d)-\b_{1}(d)+\b_{2}(d)-d/(d+1)^{2}$.
The proof is then concluded by seeing that
$\b_{0}(d)-\b_{1}(d)+\b_{2}(d)<(2d+1)/(d+1)^{2}$ for all $d\geq2$.
After some algebraic manipulations we find
$\b_{0}(d)-\b_{1}(d)+\b_{2}(d)=(d/(d-1))^{d}\frac{2(d-1)}{(d+1)(3d-1)}$.
So, we have to check that
$(d/(d-1))^{d}\frac{2(d-1)}{3d-1}<\frac{2d+1}{d+1}$ or equivalently
$\left(1-\frac{1}{d}\right)^{d}>\frac{2(d-1)(d+1)}{(3d-1)(2d+1)}$.
Developing the left term into a series and truncating it, we also
find that
$\left(1-\frac{1}{d}\right)^{d}\ge\frac{d-1}{2d}-\frac{(d-1)(d-2)}{6d^{2}}=\frac{2(d-1)(d+1)}{6d^{2}}$.
With equality only in the cases of $d=2$ and $d=3$. Furthermore, it
is clear that
$\frac{2(d-1)(d+1)}{6d^{2}}>\frac{2(d-1)(d+1)}{(3d-1)(2d+1)}$ for
$d\ge2$. So, finally, by checking the cases $d=2$ and $d=3$ directly
we conclude that the density of the second layer is strictly
dominated by the first layer density for all $d\ge2$. $\Cox$ \\


In \cite{Sudbury} the issue of comparing the behavior of the process
on a regular trees with that on a random tree having the same number
of nearest neighbors on the average was raised, and a number
of results were given. In our situation, we have the following.

\begin{theorem} Consider the random trees $S$ and $T$ with probability
generating functions $G_{S}$ and $G_{T}$ respectively. If
$G_{S}(s)>G_{T}(s)$, for all $0<s<1$ then the first layer density
of $S$ exceeds the first layer density of $T$ for all $t>0$. \\
In particular, the first layer density of the regular tree $T_{d'}$
dominates the first layer density of the regular tree $T_d$ for all
$t>0$ if $d'<d$.
\end{theorem}
\textbf{Proof:} According to Theorem \ref{theorem:randomtrees} the
density of the first layer on a random tree $S$ with probability
generating function $G(s)=\sum_{n=2}^{\infty}a_{n}s^{n}$ is given by
$\mathbb{Q}\rho_{t}^S(1) = \sum_{k=2}^{\infty} a_{k}
\frac{1-e^{-(k+1)t}}{k+1}$. Define $\gamma(t):=
\mathbb{Q}\rho_{t}^S(1) -
\mathbb{Q}\rho_{t}^T(1)= \sum_{k=2}^{\infty} (a_k - b_k) \left(\frac{1- e^{-(k+1)t}}{k+1}\right)$.%
We have $\gamma(0)=0$. In case of $G_{S}(s)>G_{T}(s)$, the time
derivative of $\gamma(t)$ becomes
$\frac{d}{dt}\gamma(t)=\sum_{k=2}^{\infty}(a_k - b_k) e^{-(k+1)t} =
e^{-t}(G_{S}(e^{-t})-G_{T}(e^{-t})) > 0$ for all $t>0$. $\Cox$

\begin{theorem}
Consider a regular tree $T_{d}$ with $d$ neighbors for each vertex,
with\\  $d \in \{ 2,3,4,\dots \}$. If random tree $S$ has an average
number of $d$ vertex neighbors, then for any $t>0$, the density of
the first layer on $S$ is higher than on $T_{d}$. \end{theorem}
\textbf{Proof:} This follows from Jensen's inequality, if we can
show that, for any fixed $t$, the function
$k\mapsto\frac{1}{k+1}(1-e^{-(k+1)t})=g_{t}(k)$ has non-negative
second derivative with respect to $k$. Indeed, a computation yields
that
$\frac{d^{2}g_{t}(k)}{dk^{2}}=\frac{2e^{-(k+1)t}}{(1+k)^{3}}\Bigl(e^{(k+1)t}-1-(k+1)t-\frac{((k+1)t)^{2}}{2}\Bigr)\geq0$.
So, $g_t(k)$ is a convex function. $\Cox$


\end{document}